\documentclass{emulateapj}
\usepackage{textcomp}
\usepackage{amsmath}
\usepackage{graphicx}
\usepackage[T1]{fontenc}

\def\beq{\begin{eqnarray}}
\def\eeq{\end{eqnarray}}

\begin{document}

\title{A method to constrain mass and spin of GRB black hole within the NDAF model}

\author{Tong Liu,\altaffilmark{1,2,3} Li Xue,\altaffilmark{1,2} Xiao-Hong Zhao,\altaffilmark{4,2} Fu-Wen Zhang,\altaffilmark{5,2} and Bing Zhang\altaffilmark{3}}

\altaffiltext{1}{Department of Astronomy, Xiamen University, Xiamen, Fujian 361005, China; lixue@xmu.edu.cn; tongliu@xmu.edu.cn}
\altaffiltext{2}{Key Laboratory for the Structure and Evolution of Celestial Objects, Chinese Academy of Sciences, Kunming, Yunnan 650011, China}
\altaffiltext{3}{Department of Physics and Astronomy, University of Nevada, Las Vegas, NV 89154, USA; zhang@physics.unlv.edu}
\altaffiltext{4}{Yunnan Observatory, Chinese Academy of Sciences, Kunming, Yunnan 650011, China}
\altaffiltext{5}{College of Science, Guilin University of Technology, Guilin, Guangxi 541004, China}

\begin{abstract}
Black holes (BHs) hide themselves behind various astronomical phenomena, and their properties, i.e., mass and spin, are usually difficult to constrain. One leading candidate for the central engine model of gamma-ray bursts (GRBs) invokes a stellar mass BH and a neutrino-dominated accretion flow (NDAF), with the relativistic jet launched due to neutrino-anti-neutrino annihilations. Such a model gives rise to a matter-dominated fireball, and is suitable to interpret GRBs with a dominant thermal component with a photospheric origin. We propose a method to constrain BH mass and spin within the framework of this model, and apply the method to a thermally-dominant GRB 101219B whose initial jet launching radius $r_0$ is constrained from the data. Using our numerical model of NDAF jets, we estimate the following constraints on the central BH: mass $M_{\rm BH} \sim 5-9~M_\odot$, spin parameter $a_* \gtrsim 0.6$, and disk mass $3~M_\odot \lesssim M_{\rm disk} \lesssim 4~M_\odot$. Our results also suggest that the NDAF model is a competitive candidate for the central engine of GRBs with a strong thermal component.
\end{abstract}


\section{Introduction}

Black holes (BHs) are mysterious and fascinating compact objects, which are sources of multi-band electromagnetic radiation, gravitational waves, neutrino emission, and cosmic rays. Two essential properties of BHs, i.e., mass and spin, are however not easy to measure. Some dynamical or statistical methods have been introduced to constrain these parameters for super-massive BHs \citep[e.g.,][]{Natarajan1998,Brenneman2006,Volonteri2007,Tchekhovskoy2010,LeiZhang2011,Kormendy2013,Wang2013} and stellar-mass BHs \citep[e.g,][]{Bahcall1978,Zhang1997,McClintock2014}.

A hyper-accreting stellar-mass BH is usually invoked as the central engine of gamma-ray bursts (GRBs)\footnote{A millisecond magnetar is another possible GRB central engine \citep{Usov92,DL98,ZM01,Metzger2011}, which likely harbors in some GRBs.}. Unlike other systems where an accretion disk or a companion is observable, the BH central engine of GRBs is completely masked by the intense emission, so that the characteristics of the BH are not easy to constrain. Nonetheless, the physics of of a neutrino-dominated accretion flow (NDAF) around the central BH has been extensively studied \citep[e.g.,][]{Popham1999,Dimatteo2002,Kohri2002,Kohri2005,Gu2006,Janiuk2007,Kawanaka2007,Liu2007,Liu2008,Liu2012a,Liu2012b,Liu2013,Liu2014,Zalamea2011,Janiuk2013b,Luo2013,Xue2013,Globus2014}. The BH properties may be inferred through confronting model predictions with the observational data.

There are two possible mechanisms to launch a relativistic jet in a hyper-accreting BH system. The first is through neutrino-anti-neutrino ($\nu \bar\nu$) annihilation from an NDAF \citep[e.g.][]{Popham1999}. Another is the Blandford-Znajek (BZ) mechanism \citep[e.g.][]{Blandford1977,Lee2000,Luo2013}. The comparisons between the emission powers of the two mechanisms have been carried out \citep{Lei2013,Liu2015b}. It is likely that the two mechanisms may play dominant roles in different parameter regimes. Observationally, some GRBs are observed to have a bright thermal spectral component that is consistent with a fireball \citep[e.g.][]{Ryde2010,Peer2012,Peer2015,Larsson2015}, even though most GRBs have no or a very weak thermal component, suggesting that the outflow may contain significant Poynting flux from the central engine \citep{Zhang2009,Gao2015}.

An NDAF is very dense and hot, and is cooled via neutrino emission. Neutrinos can tap the thermal energy gathered by the viscous dissipation and liberate a tremendous amount of binding energy, and $\nu \bar\nu$ annihilation above the disk would launch a hot fireball. A GRB powered by this mechanism is therefore thermally dominated. Within such a central engine model, the GRB luminosity depends on the mass and spin of the BH as well as the accretion rate. The launch site of the fireball, $r_0$, should be above the typical $\nu\bar\nu$ annihilation radius. Observationally, $r_0$ may be constrained by the observed thermal spectral component \citep{Peer2007}.

In this paper, we propose a method to constrain BH mass and spin of GRBs within the framework of the NDAF $\nu \bar\nu$-annihilation model. The model and the method of constraining BH parameters is presented in Section 2, and the method is applied to a thermally-dominated burst GRB 101219B in Section 3. Conclusions are presented in Section 4 with some discussion.

\section{Method}

\begin{figure}
\centering
\includegraphics[angle=0,scale=0.43]{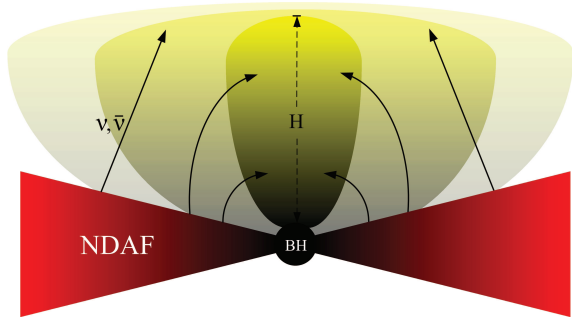}
\caption{Cartoon picture of the BH-NDAF system and its corresponding annihilation region. The red and yellow colors from light to dark of the disk and annihilation region represent the temperature and annihilation efficiency from low to high, respectively.}
\label{fig1}
\end{figure}

Within the NDAF $\nu\bar\nu$-annihilation model of GRBs, both jet luminosity ($L_{\nu\bar\nu}$) and launch radius ($r_0$) depend on the BH mass ($M_{\rm BH}$), spin ($a_*$), and also the mass accretion rate.

In our previous work \citep{Xue2013}, we investigated the relativistic one-dimensional global solutions of NDAFs by taking into account neutrino physics, balance of chemical potentials, and nucleosynthesis more precisely and in more detail than previous works. According to the 16 solutions with different characterized mass accretion rate and BH spin, we exhibited the radial distributions of various physical properties in NDAFs. We reconfirmed that electron degeneracy has an important effect, and for the first time we found that the electron fraction $Y_{\rm e}$ is about 0.46 in the outer region of the NDAF for all solutions. Furthermore, free nucleons, $\rm ^4He$, and $\rm ^{56}Fe$ are found to dominate in the inner, middle, and outer regions, respectively. We found that the neutrino trapping process can affect the value of the $\nu\bar\nu$ annihilation luminosity, especially for high accretion rate and rapid BH rotation. Finally, we approximated the neutrino luminosity, annihilation luminosity, and neutrino trapping radius with three fitting formulae as functions of BH spin and accretion rate. For general purposes, we derived the fitting formula for a very wide range of accretion rate, i.e. 0.01-10 $M_\odot~\rm s^{-1}$ \citep{Xue2013,Liu2015b}.

In order to take into account the influence of $M_{\rm BH}$, we extend our numerical solutions to explore the $M_{\rm BH}$ dependence. Meanwhile, we noticed that neutrino trapping can gradually diminish the increase of $L_{\nu\bar\nu}$ with the accretion rate $\dot M$, especially at high $\dot M$. GRB observations and theoretical considerations suggest that a relatively low accretion rate is relevant for GRBs, especially for LGRBs \citep{Liu2015b}. In the following, we only consider an accretion rate in the range of $0.01 ~M_\odot~{\rm s^{-1}} \lesssim \dot{M} \lesssim 0.5 ~M_\odot~\rm s^{-1}$ but explore the $M_{\rm BH}$ dependence.

Based on the numerical model, we derive the fitting formulae for the neutrino annihilation luminosity $L_{\nu \bar{\nu}}$ and dimensionless annihilation height $h$ as follows
\begin{eqnarray}
\log L_{\nu \bar{\nu}}  =  52.98 + 3.88  a_* - 1.55  \log m_{\rm BH} + 5.0 \log \dot{m}, \label{eq:L} \\
\log h  =  2.15 - 0.30 a_* - 0.53 \log m_{\rm BH} +  0.35 \log \dot{m}, \label{eq:H}
\end{eqnarray}
where $L_{\nu\bar\nu}$ is in units of ${\rm erg~s^{-1}}$, $a_*$ is the dimensionless spin parameter of the BH, $m_{\rm BH}=M_{\rm BH}/M_\odot$ and $\dot{m}=\dot{M}/M_\odot~\rm s^{-1}$ are the dimensionless BH mass and accretion rate, $h=H/r_{\rm g}$, $H$ is the physical annihilation height (as shown in Figure 1), and $r_{\rm g}=2GM_{\rm BH}/c^2$ is the Schwarzschild radius.

Figure 1 is a cartoon picture of the BH-NDAF system. The red color from dark to light stands for the disk temperature from high to low, and the yellow color from dark to light stands for the $\nu\bar\nu$ annihilation rate from high to low. Most of annihilation luminosity is ejected from the narrow region above the BH, and it varies much more rapidly along the vertical direction than along the radial direction of the disk according to the numerical calculations \citep[e.g.,][]{Popham1999,Liu2007,Liu2008}. \citet{Xue2013} calculated the annihilation luminosity by integrating in the entire space above the disk. Based on these results, we define the annihilation height as the region where 99.9$\%$ of the annihilation luminosity is included. The height is related to the characteristics of the BH-NDAF system as shown in Equation (2).

The analytic formula of neutrino annihilation luminosity is somewhat different from the forms in \citet{Fryer1999}\footnote{Equation (11) in \citet{Fryer1999} displays the approximate fit to the annihilation luminosity results of \citet{Popham1999}, i.e., $\log L_{\nu \bar{\nu}}  \approx  43.6 + 3.4 a_* + 4.89 \log (\dot{M}/0.01 M_\odot ~\rm s^{-1})$, for $0.01<\dot{m}<0.1$.} and \citet{Zalamea2011}. The main reasons include that (1) more detailed neutrino physics than that in \citet{Popham1999} has been considered, which can slightly affect the annihilation luminosity; that (2) a narrower range of $\dot M$ than that in \citet{Zalamea2011} has been adopted; and that (3) we have fitted the numerical results directly instead of introducing some analytical results as did in \cite{Zalamea2011}. We notice that similar fitting indices on $a_*$ and $\dot{m}$ have been also obtained by \cite{Fryer1999}, who also fitted the numerical results in the range of 0.01-0.1 $M_\odot ~\rm s^{-1}$. In a wider range of accretion rate, the neutrino annihilation luminosity increases much slower with $a_*$ and $\dot{m}$ for the accretion rates higher than, e.g., 1-10 $M_\odot ~\rm s^{-1}$, so that the fitting indices are much shallower than Equation (1) when high accretion rates are included \citep[e.g.][]{Zalamea2011,Xue2013}.

For clarification, we show $L_{\nu\bar\nu}$ dependence on $\dot m$ and $a_*$ for $M_{\rm BH} = 3 M_\odot$. The points are our numerical results, and lines are various fitting lines. One can see that in the low $\dot m$ range as assumed in this paper, the fitting line slopes are steep [solid lines correspond to Equation (1)], which are consistent with those derived by \cite{Fryer1999} (dashed lines). The dotted lines are the best fits in a much wider range of $\dot m$, so that a much shallower slope is derived \citep{Xue2013}.

From the observational viewpoint, $L_{\nu\bar{\nu}}$ may be approximated as the total jet-corrected prompt emission energy and afterglow kinetic energy divided by the duration of the burst, which can be written as \citep[e.g.,][]{Liu2015c,Song2016}
\beq L_{\nu \bar{\nu}} \approx (1+z)(E_{\rm \gamma,iso}+E_{\rm k,iso})(1-\cos \theta_{\rm j})/T_{90}, \eeq
where $E_{\rm \gamma,iso}$ and $E_{\rm k,iso}$ are the isotropic prompt $\gamma$-ray energy and kinetic energy, respectively, $z$ is the redshift, $T_{\rm 90}$ is the duration of the burst, and $\theta_j$ is the half jet opening angle. These parameters can be derived from the observational data. It should be emphasized that we assume that the jet energy is mainly due to the neutrino annihilation process rather than BH rotation energy extraction, which may not be the case for all GRBs.

The jet launching radius $r_0$ may be derived from the thermal component of a GRB, assuming that the thermal component is the emission from the fireball photosphere \citep{Meszaros2000,Peer2007}. Within the NDAF neutrino annihilation model, the annihilation height should satisfy $H \lesssim r_0$.

Finally, an accretion rate is needed in Equations (\ref{eq:L}) and (\ref{eq:H}) to derive $m_{\rm BH}$ and $a_*$. We introduce a mean dimensionless accretion rate
\beq \dot{m}\approx m_{\rm disk} (1+z)/T_{90,\rm s}, \eeq
where $m_{\rm disk}=M_{\rm disk}/M_\odot$, $M_{\rm disk}$ is the disk mass, and $T_{90,\rm s}=T_{90}/(\rm 1~s)$. Besides the constraints from Equations (1) and (2), this condition may be considered as the third condition to constrain BH mass and spin, given that the disk mass should be less than about 5 $M_\odot$ as shown from numerical simulations of collapsars \citep[e.g.,][]{Woosley1993,Popham1999,Zhang2003}. Our steady-state NDAF model in any case involves uncertainties in the accretion rate and the disk mass, which is not easy to avoid.

There is yet a fourth condition that the mass of a newborn BH from a collapsar should be larger than about 3 $M_\odot$ \citep[e.g.,][]{MacFadyen1999,Popham1999}.

For a given $\dot m = 0.1$, contours $L_{\nu\bar\nu}$ and $H$ are presented in the $m_{\rm BH} - a_*$ two dimensional plane in Figure 3 (a). The mass and spin of a BH can be constrained if $L_{\nu\bar\nu}$ and $H$ are constrained from the data based on Equations (1) and (2), given that $\dot m$ is constrained in a reasonable range based on the third and fourth conditions.

\begin{figure}
\centering
\includegraphics[angle=0,scale=0.33]{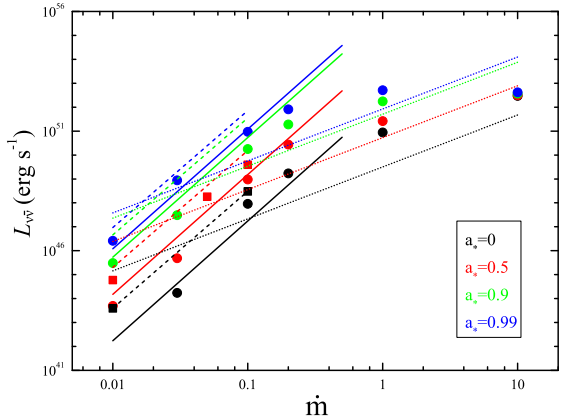}
\caption{Neutrino annihilation luminosity $L_{\nu \bar{\nu}}$ as a function of dimensionless accretion rate $\dot{m}$ for $a_*$ = 0, 0.5, 0.9, 0.99 and $M_{\rm BH}=3~M_\odot$. The circles represent our global solutions, whereas the squares represent the solutions from Table 3 in \citet{Popham1999}. The solid, dashed, and dotted lines, represent the fitting lines of Equation (1) in this paper, Equation (11) in \citet{Fryer1999}, and Equation (48) in \citet{Xue2013}, respectively.}
\label{fig2}
\end{figure}

\section{Application to GRB 101219B}

GRB 101219B, located at RA(J2000) = $00^h 49^m 02^s$, and Dec(J2000) = $-34^\circ 31' 53''$, triggered both the \emph{Swift}/BAT at 16:27:53 UT \citep{Gelbord2010} and \emph{Fermi}/GBM \citep{Horst2010}. The duration $T_{90}$ is about $51.0 ~\rm s$ in the 10-1000 keV energy band, and the redshift $z$ is about 0.55 \citep{Kienlin2014,Golkhou2015,Larsson2015}. \citet{Sparre2011} reported on the spectroscopic detection of an associated supernova, SN 2010ma, suggesting that it has a massive star core collapse origin. A significant black body component with temperature $kT=0.2 ~\rm keV$ and luminosity $\sim 10^{47} ~\rm erg~ s^{-1}$ was discovered \citep{Starling2012}. Recently, \citet{Larsson2015} analyzed the properties of its prompt emission and afterglow. Following \citet{Peer2007}, they obtained its initial Lorentz factor $\Gamma$ = $138 \pm 8$ and the jet launching radius $r_0 = 2.7 \pm 1.6 \times 10^7 ~\rm cm$, which is close to the central BH horizon radius. They also derived the isotropic energy emitted in the gamma-ray band $E_{\gamma,\rm iso} = 3.4 \pm 0.2 \times 10^{51}~ \rm ergs$, and the kinetic energy of the afterglow $E_{\rm k,iso} = 6.4 \pm 3.5 \times 10^{52}~\rm ergs$. No jet break was detected, and a lower limit on the jet opening angle, i.e. $\theta_{\rm j}>17.1^\circ$, can be inferred based on the last data point in the optical light curve. We derive the mean jet luminosity $L_{\nu\bar\nu} \gtrsim 9.0 \times 10^{49} \rm erg~s^{-1}$.

\citet{Golkhou2015} reported that the minimum variability of GRB 101219B is about $\delta t_{\rm min} = 5.386 \pm 0.868 ~\rm s$. The corresponding distance scale $c \delta t$ is much larger than $r_0$ derived by the photosphere method \citep{Larsson2015}. We therefore take $r_0$ to estimate the annihilation height $H$.

If the disk mass is set to a reasonable value (so that $\dot m$ is given), we can obtain the BH mass and spin of GRB 101219B based on the observational data. Figure 3 (b) shows the constraints by $L_{\nu\bar\nu}$ and $H$ in the $m_{\rm BH}-a_*$ plane. The solid, dashed, and dotted lines correspond to $m_{\rm disk}$=3, 3.5, and 4, respectively (which corresponds to $\dot m \approx$ 0.09, 0.12, and 0.15, respectively). We obtain ($m_{\rm BH},~a_*) \approx$ (9.37, 0.95), (6.91, 0.82), and (5.11, 0.68), respectively for $m_{\rm disk}$=3, 3.5, and 4, respectively. Notice that the calculated BH masses and spins are the average values during the burst, neither the initial nor the final values. This is because the mean jet luminosity and annihilation height of the entire burst have been used in our calculation, and because the BH properties violently evolve with time due to the high accretion rate \citep[e.g.][]{Song2015}.

In principle, $m_{\rm disk}$ is a free parameter. However, in the following we show that it is constrained in the narrow range between 3 and 4. First, if $m_{\rm disk} $ is less than 3, $a_*$ would approach 0.998, which means that the lower limit of $m_{\rm disk}$ should be about 3 (more precisely, $\sim$ 2.8). Next, we notice that the mean BH mass is about 5.11 when $m_{\rm disk}$ is set to be 4. Considering accretion process during the prompt emission phase, this may correspond to an initial mass of about 3, the possible lower limit of a nascent BH. This suggests that the $m_{\rm disk}$ cannot be much greater than 4. These values are consistent with the theoretically preferred values for a GRB central engine \citep{Woosley1993,Popham1999,Zhang2003}. In addition, since $a_* \gtrsim 0.9$ is preferred in the collapsar model \citep{Popham1999}, the preferred $m_{\rm disk}$ may be close to 3.

Based on the above results, we suggest that a Kerr BH with a mean mass $M_{\rm BH} \sim 5-9~M_\odot$ and mean spin $a_* \gtrsim 0.6$ surrounded by a disk with $3~M_\odot \lesssim M_{\rm disk} \lesssim 4~M_\odot$ may be the central engine of GRB 101219B.

\section{Conclusions and discussion}

\begin{figure}
\centering
\includegraphics[angle=0,scale=0.35]{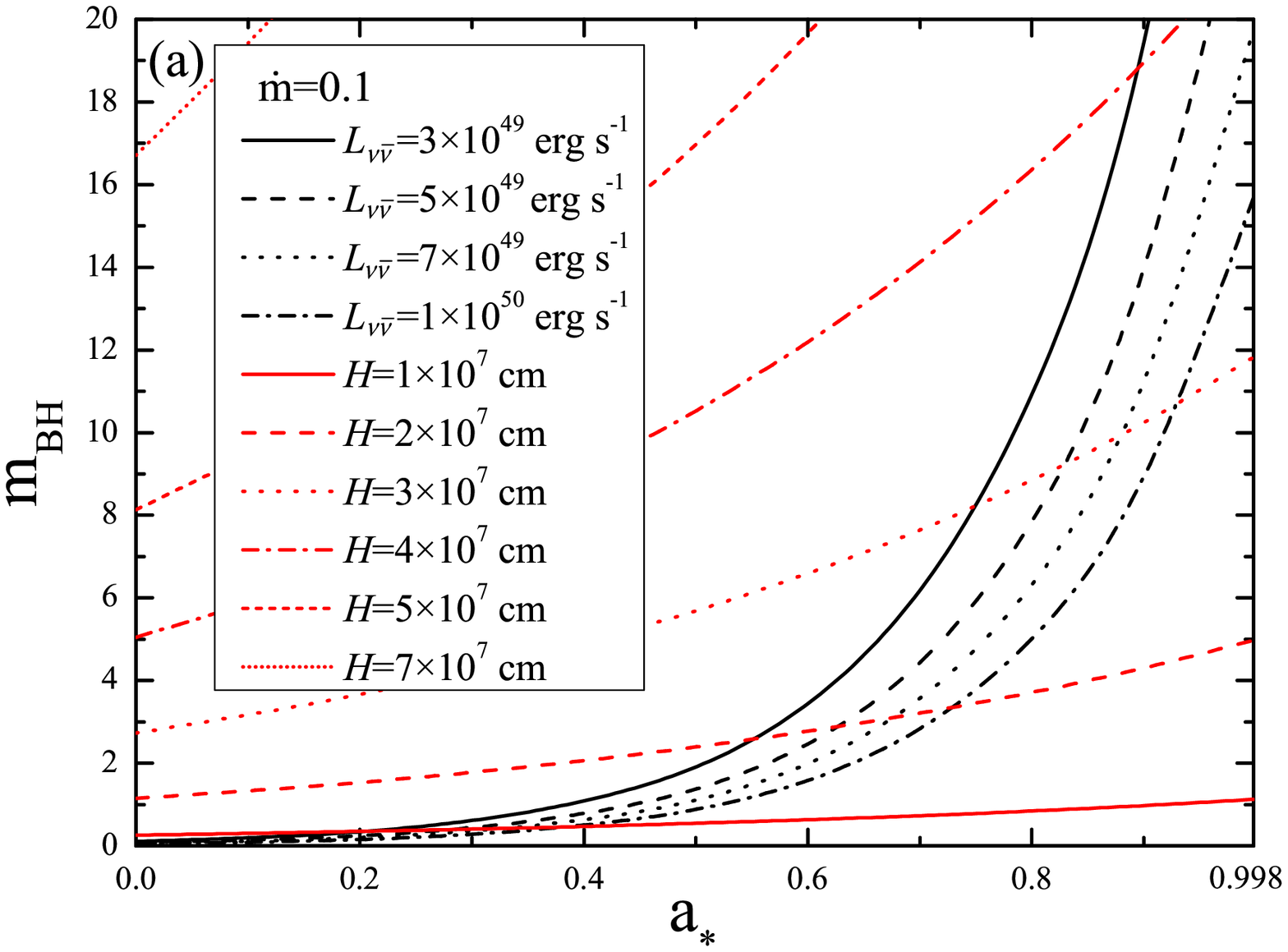}
\includegraphics[angle=0,scale=0.35]{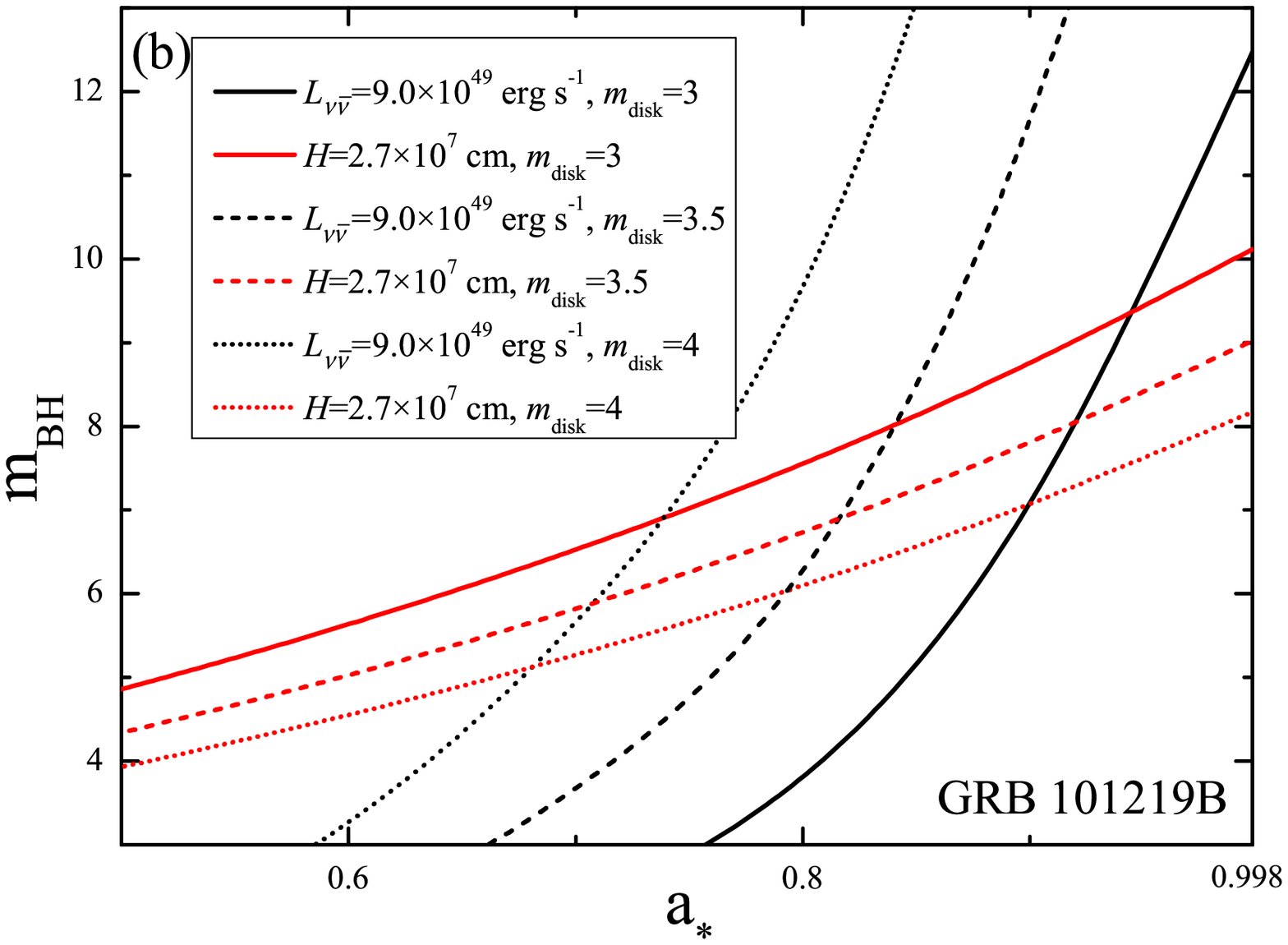}
\caption{(a) The contours of jet luminosity and annihilation height constraints on the $m_{\rm BH}-a_*$ plane based on the NDAF $\nu \bar\nu$-annihilation model. (b) The constrained mean BH mass and spin of GRB 101219B for different disk masses. The black and red lines correspond to the constraints from Equations (1) and (2), respectively. The solid, dashed, and dotted lines correspond to $m_{\rm disk}$=3, 3.5, and 4, respectively.}
\label{fig3}
\end{figure}

Within the framework of NDAF $\nu\bar\nu$-annihilation GRB central engine model, we proposed a method to constrain the mean mass and spin of the BH central engine in GRBs. The method is found applicable to the thermally dominated GRB 101219B, and the derived parameters of the BH central engine fall into the reasonable range of theoretical models.

There are several limitations of this model. First, it does not apply to those GRBs whose central engine is a magnetar. Second, within the BH central engine model, it only applies to the case that the jet is launched through the $\nu \bar\nu$-annihilation mechanism rather than magnetically-launched mechanisms such as the BZ mechanism. The method is therefore only relevant for GRBs that have a prominent thermal component, as predicted by the standard fireball model \citep{Meszaros2000}. Next, even though $L_{\nu\bar\nu}$ can be readily estimated from the prompt emission and afterglow observations, the measurement of $H$ is not straightforward. One method is to relate $H$ to $r_0$ inferred from the analysis of the thermal component of GRBs \citep{Peer2007}, as has been done for GRB 101219B in this paper. However, it is possible that in some GRBs the inferred $r_0$ is affected by the variability introduced as the jet propagates through the stellar envelope \citep{Morsony2010,Peer2015}. A possible hybrid jet composition from the central engine would also complicate the situation \citep{Gao2015}. Another method to infer $H$ may be through the observed minimum variability time scale. However, it may be limited by the count rate of the burst, and therefore may not always be a good measure of $H$. It is worth analyzing all the thermally-dominated GRBs \citep{Peer2015} using this method to check what fraction of the bursts this method is applicable, which would shed light into how dominant NDAF $\nu\bar\nu$-annihilation is in the GRB central engines. Finally, our fitting formulae [Equations (1) and (2)] are based on a linear fit to the numerical results. The uncertainties associated with such a fit would introduce an error to the constrained ranges of the parameters, which is of the same order of the ranges derived in this paper.

Besides this method, there might be other methods to constrain BH mass and spin in GRBs. First, if some GRB jets are precessing \citep[e.g.,][]{Liu2010}, the BH may capture the inner region of the NDAF to conform with the direction of the angular momentum, whereas the outer region of the disk causes the BH and inner part to precess. In this framework, the precession period or its time-evolution are related to the characteristics of the BH and disk. Combining with the observations or Equation (1), constraints on the BH mass and spin may be obtained. For example, there may exist Kerr BHs with mass $\sim 10 ~M_\odot$ in the center of GRBs 121027A and 130925A \citep{Hou2014a,Hou2014b}. Finally, the BH-NDAF precession or neutrino emission from the disk may produce violent gravitational radiation \citep[e.g.,][]{Suwa2009,Romero2010,Sun2012}. Similarly, the gravitational wave amplitudes are also connected with the characteristics of the BH accretion system, which may be used to constrain BH mass and spin.

Recently, a gravitational wave signal was observed from a binary BH system, GW 150914 \citep{Abbott2016}. This signal may be accompanied with a short GRB \citep{Connaughton2016}, and several models have been proposed to explain the observation \citep[e.g.,][]{Li2016,Zhang2016,Loeb2016,Perna2016,Liu2016,Woosley2016}. The two BH masses are determined by the signal, and this may be an independent method to measure the mass of GRB central BHs (see, e.g. an earlier investigation by \citealt{Janiuk2013a}).

\begin{acknowledgements}
We thank the anonymous referee for beneficial comments. This work is supported by the National Basic Research Program of China (973 Program) under grant 2014CB845800, the National Natural Science Foundation of China under grants 11203067, 11373002, 11473022, and U1331101, the CAS Open Research Program of Key Laboratory for the Structure and Evolution of Celestial Objects under grants OP201207 and OP201305, the Yunnan Natural Science Foundation under grants 2011FB115 and 2014FB188, and the Fundamental Research Funds for the Central Universities under grants 20720150019 and 20720160024. TL acknowledges financial support from China Scholarship Council to work at UNLV.
\end{acknowledgements}

\end{document}